\documentclass[conference]{IEEEtran}
\IEEEoverridecommandlockouts

\usepackage[hidelinks]{hyperref}

\ifCLASSOPTIONcompsoc
    \usepackage[caption=false, font=normalsize, labelfont=sf, textfont=sf]{subfig}
\else
\usepackage[caption=false, font=footnotesize]{subfig}
\fi

\usepackage{cite}
\usepackage{amsmath,amssymb,amsfonts}
\usepackage{algorithmic}
\usepackage{graphicx}
\usepackage{textcomp}
\usepackage{xcolor}

\usepackage{listings}
\usepackage{multirow}

\makeatletter
\renewcommand*\env@matrix[1][*\c@MaxMatrixCols c]{%
  \hskip -\arraycolsep
  \let\@ifnextchar\new@ifnextchar
  \array{#1}}
\makeatother

\def\BibTeX{{\rm B\kern-.05em{\sc i\kern-.025em b}\kern-.08em
    T\kern-.1667em\lower.7ex\hbox{E}\kern-.125emX}}
\begin{document}

\title{Efficiently manipulating Pauli strings with PauliArray\\
    \thanks{With the financial assistance of the government of Quebec.}
}

\author{\IEEEauthorblockN{1\textsuperscript{st} Maxime Dion}
    \IEEEauthorblockA{\textit{AlgoLab quantique, Institut quantique} \\
        \textit{Universit\'e de Sherbrooke}\\
        Sherbrooke, QC, Canada \\
        Maxime.Dion@USherbrooke.ca}
    \and
    \IEEEauthorblockN{2\textsuperscript{nd} Tania Belabbas}
    \IEEEauthorblockA{\textit{AlgoLab quantique, Institut quantique} \\
        \textit{Universit\'e de Sherbrooke}\\
        Sherbrooke, QC, Canada \\
        Tania.Belabbas@USherbrooke.ca}
    \and
    \IEEEauthorblockN{3\textsuperscript{rd} Nolan Bastien}
    \IEEEauthorblockA{\textit{D\'epartement de physique} \\
        \textit{Universit\'e de Sherbrooke}\\
        Sherbrooke, QC, Canada \\
        Nolan.Bastien@USherbrooke.ca}
}

\maketitle

\begin{abstract}

    Pauli matrices and Pauli strings are widely used in quantum computing. These mathematical objects are useful to describe or manipulate the quantum state of qubits. They offer a convenient basis to express operators and observables used in different problem instances such as molecular simulation and combinatorial optimization. Therefore, it is important to have a well-rounded, versatile and efficient tool to handle a large number of Pauli strings and operators expressed in this basis. This is the objective behind the development of the PauliArray library presented in this work. This library introduces data structures to represent arrays of Pauli strings and operators as well as various methods to modify and combine them. Built using NumPy, PauliArray offers fast operations and the ability to use broadcasting to easily carry out otherwise cumbersome manipulations. Applications to the fermion-to-qubit mapping, to the estimation of expectation values and to the computation of commutators are considered to illustrate how PauliArray can simplify some relevant tasks and accomplish them faster than current libraries.

\end{abstract}


\section{Introduction}

In quantum computing, the Pauli operators are the three single-qubit fundamental operators. Together with the identity they form a complete basis for single-qubit operators. Joining $n$ of them using the tensor product produces a $n$-qubit operator called a Pauli string. The set of $4^n$ Pauli strings is therefore a complete basis for $n$-qubit operators.

While these multi-qubit operators are widely used in the context of stabilizer codes\cite{gottesman_stabilizer_1997,aaronson_improved_2004}, their use has been spreading to other aspects of quantum computing. For instance, linear combinations of Pauli strings are used to represent observables such as Hamiltonians in various variational quantum algorithms, including the variational quantum eigensolver (VQE) \cite{peruzzo_variational_2014,bharti_noisy_2022} and the quantum approximate optimization algorithm (QAOA) \cite{farhi_quantum_2019}. This representation allows for the efficient estimation of the expectation value through qubit measurements. Moreover, representing a Hamiltonian in this way is useful for simulating the evolution of a quantum state by Trotterization \cite{lloyd_universal_1996}.

These applications require handling numerous Pauli strings through combining them into products, commutators, linear combinations and other convoluted manipulations. It is therefore important to have well-rounded, versatile and efficient software tools to represent and manipulate these objects. Indeed, all major quantum computing libraries such as Qiskit\cite{qiskit_contributors_qiskit_2023}, Pennylane\cite{bergholm_pennylane_2022} and Cirq\cite{developers_cirq_2023} implement data structures for that purpose.

The present work distinguishes itself from current libraries by introducing the notion of multidimensional arrays of Pauli strings. This multidimensionality is often implied in mathematical expressions but the current tools cannot handle it directly. Therefore, the library PauliArray aims to provide all basic functionalities offered in major quantum computing libraries while improving them by offering the convenience of multidimensionality and the flexibility of broadcasting. The objective of this paper is to introduce the data structures of the library, their main functionalities, and how they can simplify otherwise cumbersome operations. To learn how to use this library, please refer to the documentation of PauliArray~\cite{algolab_quantique_pauliarray_2024}.

Before introducing the library, the properties of Pauli strings are briefly reviewed in section~\ref{sec:pauli_strings}. The main data structures and their functionalities are introduced in section~\ref{sec:data_structures} and~\ref{sec:functionnalities}. To illustrate how multidimensional arrays of Pauli strings and operators can be useful, section~\ref{sec:use_cases} demonstrates how PauliArray is used in the contexts of the Hamiltonian preparation with the fermion-to-qubit mapping, the computation of commutators for the Adaptative Derivative Assembled Pseudo-Trotter algorithm (ADAPT) and the estimation of expectation values of observables.

\section{Pauli Strings}
\label{sec:pauli_strings}

Every operator acting on a single qubit can be expressed as a linear combination of the identity $\hat{I}$ and the three Pauli operators $\hat{X}$, $\hat{Y}$ and $\hat{Z}$. In the computational basis, these operators can respectively be represented by the following matrices
\begin{align*}
    \mathsf{I} & = \begin{pmatrix}
                       1 & 0 \\ 0 & 1
                   \end{pmatrix}, &  &            &
    \mathsf{X} & = \begin{pmatrix}
                       0 & 1 \\ 1 & 0
                   \end{pmatrix},                   \\
    \mathsf{Y} & = \begin{pmatrix}
                       0 & -i \\ i & 0
                   \end{pmatrix} &  & \text{and} &
    \mathsf{Z} & = \begin{pmatrix}
                       1 & 0 \\ 0 & -1
                   \end{pmatrix}.
\end{align*}
Each of these four operators can be mapped to two binary variables $z$ and $x$ ($\in\{0,1\}$). The operator can be reconstructed using the equation $(-i)^{zx}\hat{Z}^{z} \hat{X}^{x}$. Table~\ref{table:pauli_bits} shows the mapping between the binary variables and the operators.

\subsection{Definitions and Notations}

A Pauli string of length $n$ is an operator acting on $n$ qubits defined as the tensor product
\begin{align*}
    \hat{P}        & = \bigotimes_{q=0}^{n-1} \hat{\sigma}_q
                   &                                                    & \text{where} &
    \hat{\sigma}_q & \in \left\{\hat{X},\hat{Y},\hat{Z},\hat{I}\right\}
\end{align*}
is one of the Pauli operators, or the identity operator, acting on qubit $q$. To encode a Pauli string it is sufficient to provide two vectors of $n$ binary components, also called bit strings, $\mathbf{z} = (z_0, z_1, \ldots, z_{n-1})$ and $\mathbf{x} = (x_0, x_1, \ldots, x_{n-1})$. The Pauli string can be constructed using
\begin{align}
    \hat{P}
     & = (-i)^{\mathbf{z} \cdot \mathbf{x}} \hat{Z}^{\mathbf{z}} \hat{X}^{\mathbf{x}}
    \label{eq:one_pauli}
\end{align}
where the exponentiation by a vector is to be interpreted as a tensor product
\begin{align*}
    \hat{Z}^{\mathbf{z}} \equiv  \bigotimes_{q=0}^{n-1} \hat{Z}^{z_q}.
\end{align*}
Some shortened notations are used in this paper. From now on, whenever an index is repeated between a Pauli operator and its exponent, a tensor product is assumed
\begin{align*}
    \hat{Z}_q^{z_q} \equiv  \bigotimes_{q=0}^{n-1} \hat{Z}^{z_q}.
\end{align*}
In section~\ref{ssec:fermions_qubits_map}, the vector exponent will be the result of a matrix multiplication. The following repeated indices notation will also be used
\begin{align*}
    \hat{Z}_q^{A_{qp}b_p} \equiv \bigotimes_{q=0}^{n-1} \hat{Z}^{\sum_{p=0}^{n-1} A_{qp}b_p}.
\end{align*}

\subsection{Operations on Pauli Strings}

The composition of two Pauli strings produces another Pauli string up to a phase factor
\begin{align}
    \hat{P}_1 \hat{P}_2 = (-i)^{f}\hat{P}_3
    \label{eq:pauli_composition}
\end{align}
where the resulting Pauli string is described by the two bit strings
\begin{align*}
    \mathbf{x}_3 = \mathbf{x}_1 + \mathbf{x}_2 \pmod{2}
    \quad\text{and}\quad
    \mathbf{z}_3 = \mathbf{z}_1 + \mathbf{z}_2 \pmod{2}
\end{align*}
and where the phase factor is given by
\begin{align*}
    f & =
    2\mathbf{x}_1 \cdot \mathbf{z}_2
    + \mathbf{z}_1 \cdot \mathbf{x}_1
    + \mathbf{z}_2 \cdot \mathbf{x}_2
    - \mathbf{z}_3 \cdot \mathbf{x}_3 \pmod{4}.
\end{align*}

The composition of Pauli strings allows the computation of commutators. Two Pauli strings either commute or anticommute. Therefore,
\begin{align}
    [\hat{P}_1, \hat{P}_2] =
    \begin{cases}
        0                     & \text{if they commute,} \\
        2 \hat{P}_1 \hat{P}_2 & \text{otherwise.}
    \end{cases}
    \label{eq:commutator}
\end{align}
It is, however, not necessary to compute their commutator to verify of two Pauli strings commute. Indeed, by simply using \eqref{eq:one_pauli}, it is easy to show that
\begin{align}
    \hat{P}_1\hat{P}_2 = (-1)^{\mathbf{z}_1 \cdot \mathbf{x}_2 + \mathbf{x}_1 \cdot \mathbf{z}_2} \hat{P}_2\hat{P}_1.
    \label{eq:commutation_rule}
\end{align}
Therefore, two Pauli strings commute iff $\mathbf{z}_1 \cdot \mathbf{x}_2 + \mathbf{x}_1 \cdot \mathbf{z}_2 = 0 \pmod{2}$, and anticommute otherwise.

\begin{table}
    \centering
    \caption{Binary encoding of the three Pauli operators and the identity into two bits.}
    \begin{tabular}{c|c||c}
        $z$ & $x$ & $(-i)^{zx}\hat{Z}^{z} \hat{X}^{x} $ \\\hline
        0   & 0   & $\hat{I}$                           \\
        1   & 0   & $\hat{Z}$                           \\
        0   & 1   & $\hat{X}$                           \\
        1   & 1   & $\hat{Y}$
    \end{tabular}
    \label{table:pauli_bits}
\end{table}

\section{Data Structures}
\label{sec:data_structures}

\begin{figure*}
    \centering
    \hfill
    \subfloat[Bit string array\label{subfig:bit_array}]{\includegraphics{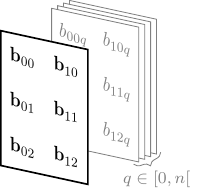}} \hfill
    \subfloat[Pauli array\label{subfig:pauli_array}]{\includegraphics{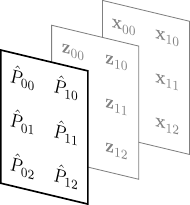}} \hfill
    \subfloat[Weighted Pauli array\label{subfig:weighted_pauli_array}]{\includegraphics{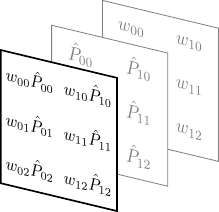}} \hfill{ } \\
    \hfill
    \subfloat[Operator\label{subfig:operator}]{\includegraphics{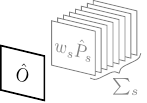}} \hfill
    \subfloat[Operator array (type 1)\label{subfig:operator_array_type_1}]{\includegraphics{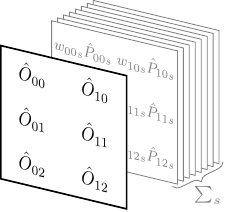}} \hfill
    \subfloat[Operator array (type 2)\label{subfig:operator_array_type_2}]{\includegraphics{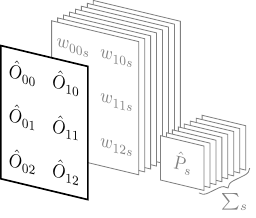}} \hfill{ }
    \caption{The data structures of PauliArray.}
    \label{fig:data_structures}
\end{figure*}

In PauliArray, a bit string is stored as a vector $\mathbf{b}$ of binary components (mod 2 integers). Placing many of these bit strings within a $d$-dimension array is possible by using a $(d+1)$-dimension bit array $\mathsf{b}$, as shown on figure~\ref{subfig:bit_array} where the last \emph{hidden} dimension is along the length of the Pauli strings and is of size of $n$. The elements of these arrays are related to the bits of the bit string such that
\begin{align*}
    [\mathsf{b}]_{ij\ldots k q} = [\mathbf{b}_{ij\ldots k}]_q
\end{align*}
where $\mathbf{b}_{ij\ldots k}$ is the $ij\ldots k$th element of the bit string array and $q$ is the qubit index. The multi-dimensional array data structure from the library NumPy\cite{harris_array_2020} is used as the backend to store these arrays of binary values as booleans.

\subsection{Array of Pauli Strings}

The basic data structure of PauliArray library represents a $d$-dimension array of Pauli strings and is called a PauliArray. As shown in figure~\ref{subfig:pauli_array}, it uses two arrays of the same shape of bit strings $\mathbf{x}$ and $\mathbf{z}$ to store this information such that each Pauli string in the PauliArray is given by
\begin{align}
    \hat{P}_{ij\ldots k} & = (-i)^{\mathbf{z}_{ij\ldots k} \cdot \mathbf{x}_{ij\ldots k}} \hat{Z}^{\mathbf{z}_{ij\ldots k}} \hat{X}^{\mathbf{x}_{ij\ldots k}}
    .
    \label{eq:one_pauli_array}
\end{align}
For example, three two-qubit Pauli strings can be defined by two $3\times 2$ bit arrays
\begin{align*}
    \mathsf{x} = \begin{pmatrix}
                     0 & 0 \\ 1 & 0 \\ 1 & 1
                 \end{pmatrix}
    ,
    \mathsf{z} = \begin{pmatrix}
                     0 & 1 \\ 0 & 0 \\ 1 & 0
                 \end{pmatrix}
    \to
    \begin{pmatrix}
        \hat{Z}\hat{I} \\ \hat{I}\hat{X} \\ \hat{X}\hat{Y}
    \end{pmatrix}
\end{align*}
where the Pauli string labels are written using the little-endian convention as they will be throughout this paper.

A PauliArray can be instantiated by providing two bit arrays $\mathsf{x}$ and $\mathsf{z}$ of the same shape. A more convenient initialization method using Pauli string labels is also available.





\subsection{Array of Weighted Pauli Strings}

A WeightedPauliArray (figure~\ref{subfig:weighted_pauli_array}) is obtained by assigning a complex number to each Pauli string in a PauliArray
\begin{align}
    w_{ij\ldots k} \hat{P}_{ij\ldots k} .
    \label{eq:one_weighted_pauli_array}
\end{align}
It can be initialized by providing a PauliArray and an array of complex coefficients of the same shape.


\subsection{Operator}

Any $n$ qubits operator $\hat{O}$ can be decomposed on the basis of Pauli strings of length $n$
\begin{align}
    \hat{O} = \sum_s w_s \hat{P}_s.
    \label{eq:one_operator}
\end{align}
Therefore an Operator, as shown on figure~\ref{subfig:operator}, is simply a sum over a one-dimensional WeightedPauliArray. It can be initialized by simply providing a one-dimensional WeightedPauliArray.


\subsection{Array of Operators (type 1)}

Using the same logic, it is possible to define an array of operators by simply using the data structure of WeightedPauliArray and using its last dimension as the summation axis
\begin{align}
    \hat{O}_{ij\ldots k} = \sum_s w_{ij\ldots ks} \hat{P}_{ij\ldots ks}.
    \label{eq:one_operator_array_type_1}
\end{align}
This type of operator array is quoted as type 1 and is illustrated on figure~\ref{subfig:operator_array_type_1}. All the operators in it have the same number of Pauli strings. It is, however, possible to set some coefficients to zero to have operators with different numbers of terms. In this case, the last dimension of the weight array is equal to the largest number of terms. This structure is most suitable when all the operators in the array decompose on disjoint sets of Pauli strings.


\subsection{Array of Operators (type 2)}

For the second type of operator array, all the operators are decomposed on the same basis of Pauli strings
\begin{align}
    \hat{O}_{ij\ldots k} = \sum_s w_{ij\ldots ks} \hat{P}_{s}.
    \label{eq:one_operator_array_type_2}
\end{align}
This type of operator array is useful when many Pauli strings are shared between the different operators. As shown on figure~\ref{subfig:operator_array_type_2}, to initialize such an operator array, one needs to provide an array of weights and a basis of Pauli strings in the form of a one-dimensional PauliArray.

\subsection{Common Attributes}

For all data structures, the number of qubits is an attribute. All array-like data structures have a number of dimensions and a shape.

\subsection{Common Methods}

Some methods are shared between all or some data structures. For instance, the Hermitian adjoint can be obtained for any data structure. Since Pauli strings are Hermitian, this operation simply takes the complex conjugate of the associated weights.

Array-like data structures offer common methods to reshape or flatten to modify its shape. These methods only act on the shape of the data structure and do not interfere with the \emph{hidden} dimension of the underlying bit string arrays or the summation dimension for operator arrays.

The \emph{unique} method can be used to extract the unique Pauli strings from any data structure. This method is useful to recombine the coefficients of an operator associated to the same Pauli strings.

Finally, all data structures offer an \emph{inspect} method which produces a description of its data.

\section{Functionalities}
\label{sec:functionnalities}

All the data structures of PauliArray can be acted on or combined using a set of operations. Some of these operations are common and shared between all the data structures. Throughout this section, we will assume one-dimensional arrays to lighten the notation, but these operations apply to multidimensional arrays unless stated otherwise.

Most of the operations described here only make sense for Pauli strings of the same length. Also, while it is possible to infer expected behaviour for operations between different kinds of data structures, at the moment most operations involving two objects can only be performed between objects of the same data structure. This might change in the future.

Like in NumPy, the operations in PauliArray between array-like data structures are done element-wise. This means these operations can only be applied on arrays of the same shape or arrays that are broadcastable following NumPy rules on broadcasting \cite{harris_array_2020}. Broadcasting is one of the features that make PauliArray a flexible tool.



\begin{figure}
    \centering
    \subfloat[Element-wise composition\label{subfig:element_wise_compose}]{\includegraphics{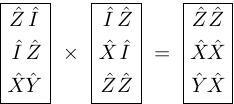}}
    \\
    \subfloat[Composition using broadcasting\label{subfig:broadcast_compose}]{\includegraphics{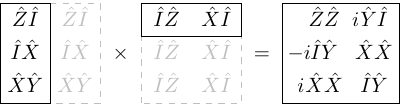}}
    \caption{PauliArray operations such as composition are done element-wise. The use of broadcasting enables to carry out more complex operations.}
    \label{fig:key_operations}
\end{figure}

\subsection{Composition}

The operation of acting on an operator with another operator is called composition. It is equivalent to a matrix product between the matrices representation of the operators. In PauliArray the composition of two arrays is element-wise as shown on figure~\ref{subfig:element_wise_compose}. Following \eqref{eq:pauli_composition}, the element-wise composition of two WeightedPauliArray of the same shape produces a third one
\begin{align*}
    w_i^{(1)}\hat{P}_i^{(1)} w_i^{(2)} \hat{P}_i^{(2)} = w_i^{(3)} \hat{P}_i^{(3)}
\end{align*}
with weights $w_i^{(3)} = (-i)^{f_i} w_i^{(1)} w_i^{(2)}$.

However, with broadcasting it is possible to perform composition in a outer product fashion such as
\begin{align*}
    w_i^{(1)}\hat{P}_i^{(1)} w_j^{(2)} \hat{P}_j^{(2)} = w_{ij}^{(3)} \hat{P}_{ij}^{(3)}.
\end{align*}
Figure~\ref{subfig:broadcast_compose} illustrates this with an example. This is how PauliArray handles the composition between two operators
\begin{align*}
    \hat{O}^{(1)} \hat{O}^{(2)} = \sum_{i,j} w_i^{(1)} \hat{P}_i^{(1)} w_j^{(2)} \hat{P}_j^{(2)}
    = \sum_k w_{k}^{(3)} \hat{P}_{k}^{(3)}
\end{align*}
where the sum over $k$ is a simple reshape over the sums over $i$ and $j$. Similar approaches are used for the composition of Operator arrays and are also compatible with broadcasting.

\subsection{Clifford Conjugation}

A set of operations very specific to Pauli strings are Clifford operations. The conjugation of a Pauli string by a Clifford operation $\hat{C}$ return another Pauli string with a possible sign change.

In PauliArray, all the classes can be transformed using built-in Clifford conjugation with operations such as $\hat{X}$, $\hat{H}$, $\hat{S}$, $C\hat{X}$ and $C\hat{Z}$. The result is a data structure where the Pauli strings have been transformed element-wise
\begin{align*}
    \hat{C}\hat{P}_i\hat{C}^\dagger = \pm\hat{P}_i^\prime.
\end{align*}

This is done by calling the appropriate method and by specifying the qubits to apply the operation on. For efficiency, these methods acts directly on the $\mathbf{x}$ and $\mathbf{z}$ bit string arrays using specific rules for each Clifford operation\cite{berg_circuit_2020}.

One can also act with a custom Clifford operator by providing an appropriate Operator. In this case, composition and simplification are used to compute the resulting Pauli strings.

\subsection{Commutation}
\label{ssec:commutation}

In PauliArray, all the commutator computations rely on the element-wise commutators between two arrays of Pauli strings
\begin{align*}
    [\hat{P}^{(1)}_i, \hat{P}^{(2)}_i] = \hat{P}^{(1)}_i \hat{P}^{(2)}_i - \hat{P}^{(2)}_i \hat{P}^{(1)}_i
    .
\end{align*}
This operation can be simplified by relying on~\eqref{eq:commutator} and the commutation rule~\eqref{eq:commutation_rule} to reduce it to a single composition
\begin{align*}
    [\hat{P}^{(1)}_i, \hat{P}^{(2)}_i] = 2c_i \hat{P}^{(1)}_i \hat{P}^{(2)}_i
\end{align*}
where
\begin{align*}
    c_{i} = \mathbf{z}^{(1)}_i \cdot \mathbf{x}^{(2)}_i + \mathbf{x}^{(1)}_i \cdot \mathbf{z}^{(2)}_i \pmod{2}
\end{align*}
is $1$ if $[\hat{P}^{(1)}_i, \hat{P}^{(2)}_j] \neq 0$.

The commutator between two operators is handled a bit differently. The general expression for such a commutator is
\begin{align*}
    [\hat{O}^{(1)}, \hat{O}^{(2)}] = \sum_{i,j} w_i^{(1)} w_j^{(2)}  [\hat{P}_i^{(1)}, \hat{P}_j^{(2)}]
\end{align*}
where $[\hat{P}_i^{(1)}, \hat{P}_j^{(2)}]$ could be evaluated using broadcasting. However, all the commuting pairs of Pauli strings will not contribute to this commutator. It is therefore more efficient, to first use the commutation rule~\eqref{eq:commutation_rule}
\begin{align*}
    c_{ij} = \mathbf{z}^{(1)}_i \cdot \mathbf{x}^{(2)}_j + \mathbf{x}^{(1)}_i \cdot \mathbf{z}^{(2)}_j \pmod{2}
\end{align*}
to identify the indices of non-commuting elements
\begin{align*}
    I_\mu, J_\mu \quad \text{for} \quad c_{I_\mu J_\mu} = 1.
\end{align*}
The commutator can then be expressed as a single sum
\begin{align*}
    [\hat{O}^{(1)}, \hat{O}^{(2)}] = 2\sum_{\mu}  c_{I_\mu J_\mu}  w_{I_\mu}^{(1)} w_{J_\mu}^{(2)} \, \hat{P}_{I_\mu}^{(1)} \hat{P}_{J_\mu}^{(2)}.
\end{align*}
This has the advantage that the number of Pauli strings to compose can be greatly reduced. It also reduces the number of repeated Pauli strings in the sum which need to be recombined to express the operator on a basis on unique Pauli strings.

\subsection{Multiplication}

All classes having weights can be multiplied by an array of numbers (or a single number) with the condition that their shapes are broadcastables
\begin{align*}
    c_{i} w_{i} \hat{\mathcal{P}}_{i} = w_{i}^\prime \hat{\mathcal{P}}_{i}
\end{align*}
with the new weights $w_{i}^\prime = c_{i} w_{i}$.

The multiplication of a PauliArray by a number or an array of numbers will produce a WeightedPauliArray.

\subsection{Addition}

Operators can be added to one another to produce a new operator. The same is true for operator arrays of the same data structure and broadcastable shapes. In this case the addition is done element-wise.

The addition between instances of other data structures (PauliArray and WeightedPauliArray) of broadcastable shapes is also possible and yields type 1 operator arrays.

\subsection{Summation}

All array-like data structures can be summed using the summation method and by providing an axis argument to specify in which directions to perform the operation. The result of this operation is always a type 1 operator array unless it contains a single operator, in which case it returns an Operator.

\subsection{Simplification}
\label{ssec:simplification}

When an Operator is the result of a composition or an addition, it is possible that a single Pauli strings appears multiple times in the summation. Whenever this occurs, these terms can be combined in a single one by adding their associated weights
\begin{align*}
    w_1\hat{P} + w_2\hat{P} = (w_1 + w_2)\hat{P}
\end{align*}
and therefore reducing the number of terms in the summation. Following this, some weights might be equal to zero or very small. Given a certain threshold value, all the terms with smaller weights can be ignored and removed from the sum, reducing again the total number of terms. Applying these two processes to an operator is called simplification.

\subsection{Tensor Product}

The tensor product of two Pauli strings of length $n_1$ and $n_2$ will produce a Pauli string of length $n_1 + n_2$. Therefore, the tensor product of two WeightedPauliArray will produce a third one
\begin{align*}
    (w_i^{(1)}\hat{P}_i^{(1)} ) \otimes (w_i^{(2)} \hat{P}_i^{(2)}) = w_i^{(3)} \hat{P}_i^{(3)}
\end{align*}
with $w_i^{(3)} = w_i^{(1)} w_i^{(2)}$ and  $\hat{\mathcal{P}}_{i}^{(3)} = \hat{\mathcal{P}}_{i}^{(1)} \otimes \hat{\mathcal{P}}_{i}^{(2)}$. Naturally, this method works for data structures with different numbers of qubits.

Again, this operation is compatible with broadcasting which enables to perform tensor product between Operators and Operator arrays.


\subsection{Indexing and Masking}

Elements of array-like data structures in PauliArray can be accessed using indexing, slicing and masking in the same manner as in NumPy. The new axis approach is particularly useful for broadcasting.

To emphasize on the array nature of the PauliArray data strutures, the hidden dimensions along the bit strings or along the summation for arrays of operators are not accessible. To extract Pauli strings for a subset of qubits, special methods must be used.

\subsection{Conversion}

With interoperability and compatibility in mind, PauliArray provides a conversion paradigm that allows for integration with other libraries, such as Qiskit, PennyLane, and OpenFermion, transforming PauliArray objects into equivalent data structures and vice versa. However, due to PauliArray's multidimensional nature, multidimensional arrays need to be flattened before they can be exported.

\section{Use cases}
\label{sec:use_cases}

PauliArray is aimed at being a flexible and general tool to manipulate Pauli strings. The present section illustrates its qualities by describing how it can be applied to well-known procedures.

\subsection{Fermion-to-qubit Mapping}
\label{ssec:fermions_qubits_map}

The mapping from fermionic occupational state on $n$ orbitals to $n$-qubit basis states
\begin{align*}
    |f_0,f_1,\ldots,f_{n-1}\rangle \to |b_0,b_1,\ldots,b_{n-1}\rangle
\end{align*}
can be defined in a general way where the state of each qubit in a basis state depends on the occupations of the orbitals following
\begin{align*}
    b_q = \sum_{p=0}^{n-1} M_{qp} f_p \pmod{2}
\end{align*}
where $M_{qp}$ are binary matrix elements of a $n\times n$ mapping matrix $\mathsf{M}$. The mapping matrix has to be invertible in the binary field so that
\begin{align*}
    \sum_p [\mathsf{M}^{-1}]_{rq} M_{qp} = \delta_{rq} \pmod{2}.
\end{align*}
For example, using the identity matrix for $\mathsf{M}$ leads to the Jordan-Wigner mapping.

The choice of a mapping matrix also defines how to translate fermionic creation and annihilation operators to qubits operators in the form of linear combinations of Pauli strings. For a general mapping \cite{steudtner_fermion--qubit_2018}, each creation/annihilation operator ($+/-$) can be written as
\begin{align}
    \hat{a}_i^\pm & =
    \frac{1}{2}
    \hat{X}_q^{M_{qi}}
    \hat{Z}_q^{\theta_{ik}[\mathsf{M}^{-1}]_{kq}}
    \left(1 \pm \hat{Z}_q^{ [\mathsf{M}^{-1}]_{iq} }\right)
    \label{eq:creation_annihilation_operators}
\end{align}
where $[\mathsf{M}^{-1}]_{ij}$ is a matrix element of the inverse of the mapping matrix and
\begin{align}
    \theta_{ij} =
    \begin{cases}
        0 & i \leq j \\
        1 & i > j
    \end{cases}
    \label{eq:heavyside_matrix}
\end{align}
is a discrete version of the Heavyside function.

Ultimately, the mapping converts a fermionic Hamiltonian
\begin{align*}
    \hat{H} = \sum_{i,j=0}^{n-1} h_{ij}\, \hat{a}^\dagger_i \hat{a}_j + \frac{1}{2} \sum_{i,j,k,l=0}^{n-1} g_{ijkl}\, \hat{a}^\dagger_i \hat{a}^\dagger_j \hat{a}_k \hat{a}_l
\end{align*}
into a linear combination of Pauli strings
\begin{align*}
    \hat{H} = \sum_{i} h_i \hat{P}_i.
\end{align*}

PauliArray is a convenient tool to handle the numerous Pauli strings involved in a fermion-to-qubit mapping. It offers ready to use implementations of this kind of mapping within the mapping submodule.

To initialize a generic mapping, it suffices to provide a mapping matrix respecting the requirement stated earlier. Predefined mapping such as Jordan-Wigner, Parity and Bravyi-Kitaev are also available and can be initialized by simply providing a number of orbitals/qubits $n$. Once a mapping is defined, PauliArray offers two main approaches to carry it out.

\subsubsection{Dense Hamiltonian}

The first approach consists of constructing the creation and the annihilation operators into two one-dimensional type 1 operator arrays using~\eqref{eq:creation_annihilation_operators}. Then, broadcasting and composition can be used to obtain the operator arrays representing
\begin{align*}
    \hat{a}^\dagger_i \hat{a}_j = \sum_{s=1}^{4} w_{ij s} \hat{P}_{ij s}
\end{align*}
and
\begin{align*}
    \hat{a}^\dagger_i \hat{a}^\dagger_j \hat{a}_k \hat{a}_l = \sum_{s=1}^{16} w_{ijkl s} \hat{P}_{ijkl s}
    .
\end{align*}

These operator arrays can be obtained directly from a mapping instance and then be multiplied by the number arrays for $g_{ij}$ and $g_{ijkl}$. Finally, the summation can be performed over all indices to get a single Operator which can finally be simplified.

This method is suitable when the coefficient arrays are dense since all operators are constructed regardless if their associated coefficients are zero or not.

\subsubsection{Sparse Hamiltonian}

The second approach aims to optimize the construction of the Hamiltonian for sparse coefficient arrays. To achieve this,
sparse coefficient arrays need to be translated into sparse representation using the coordinate format (COO). The one-body coefficient array is translated to a list of coefficients and two lists of indices
\begin{align*}
    h_{ij} \to \tilde{h}_{\mu}, I_\mu, J_\mu \quad\text{for}\quad h_{ij} \neq 0
\end{align*}
so that $h_{I_\mu J_\mu} = \tilde{h}_\mu \neq 0$. In a similar way, the two-body coefficient array is translated to a list of coefficients and four lists of indices
\begin{align*}
    g_{ijkl} \to \tilde{g}_{\nu}, I_\nu, J_\nu, K_\nu, L_\nu \quad\text{for}\quad g_{ijkl} \neq 0
\end{align*}
so that $g_{I_\nu J_\nu K_\nu L_\nu} = \tilde{g}_\nu \neq 0$.

Copies of \eqref{eq:creation_annihilation_operators} with different indices can be composed to directly express the one-body and the two-body operators. The appendix shows how the one-body operator can be written as a combination of three operators
\begin{align}
    \hat{a}_{i}^\dagger \hat{a}_{j}
     & =
    \frac{1}{4}
    \hat{U}^{(2)}_{ij} \hat{F}^{(2+)}_{ij} \hat{F}^{(1-)}_{j}
    \label{eq:one_body_operator_paulis}
\end{align}
and the two-body operator as the combination of five operators
\begin{align}
    \begin{split}
        \hat{a}_i^\dagger \hat{a}_j^\dagger \hat{a}_k \hat{a}_l
         & =
        \frac{1}{16} \,
        \hat{U}^{(4)}_{ijkl} \hat{F}^{(4+)}_{ijkl} \hat{F}^{(3+)}_{jkl} \hat{F}^{(2-)}_{kl} \hat{F}^{(1-)}_{l}.
    \end{split}
    \label{eq:two_body_operator_paulis}
\end{align}
All the operators are defined in the appendix in~\eqref{eq:update_flip_op_arrays_one} and~\eqref{eq:update_flip_op_arrays_two}.

These operators are required only for indices associated with non-zeros elements in the coefficient array. Using the list of indices, the required elements for one-body and two-body operators can be translated into one-dimensional type 1 operator arrays
\begin{align*}
    \hat{U}^{(4)}_{\nu} & = \hat{U}^{(4)}_{I_\nu J_\nu K_\nu L_\nu}
    \\
    \hat{U}^{(2)}_{\mu} & = \hat{U}^{(2)}_{I_\mu J_\mu}
    \\
                        & \ldots
\end{align*}

Finally, the various operator arrays are combined element-wise and using broadcasting following~\eqref{eq:one_body_operator_paulis} and~\eqref{eq:two_body_operator_paulis} and then summed over $\mu$ and $\nu$ respectively to get the one-body and two-body Hamiltonians
\begin{align*}
    \sum_{ij} h_{ij}\, \hat{a}_{i}^\dagger \hat{a}_{j}
    = \frac{1}{4} \sum_\mu \tilde{h}_\mu\,
    \hat{U}^{(2)}_\mu \hat{F}^{(2+)}_\mu\hat{F}^{(1-)}_\mu
\end{align*}
and
\begin{align*}
    \sum_{ijkl} g_{ijkl}\, \hat{a}_{i}^\dagger \hat{a}_{j}^\dagger \hat{a}_{k} \hat{a}_{l}
    =
    \frac{1}{16} \sum_\mu \tilde{g}_\nu
    \,
    \hat{U}^{(4)}_{\nu} \hat{F}^{(4+)}_{\nu} \hat{F}^{(3+)}_{\nu} \hat{F}^{(2-)}_{\nu} \hat{F}^{(1-)}_{\nu}
    .
\end{align*}
All of these steps are handled by the mapping instance.

\begin{figure}
    \centering
    \includegraphics{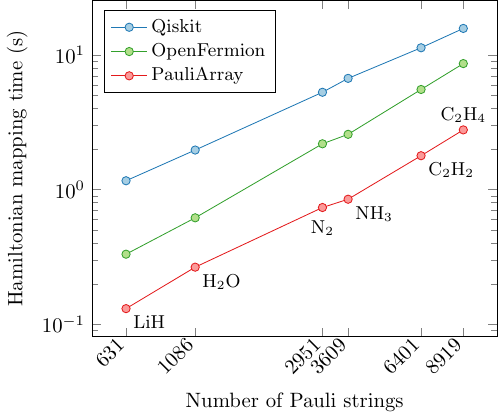}
    \caption{Average time to perform the Jordan-Wigner mapping for various small molecules with Qiskit, OpenFermion and PauliArray.}
    \label{fig:hamiltonian_mapping_plot}
\end{figure}

To illustrate that PauliArray can complete all these steps efficiently, it was used to perform the Jordan-Wigner mapping of the Hamiltonian for a selection of small molecules listed on table~\ref{table:pauli_bits} with their respective numbers of orbitals (number of qubits) and the number of Pauli strings in the resulting Hamiltonian. Since such Hamiltonians are sparse, the second approach was used. For comparison, the time required to perform the mappings with the implementations of Qiskit\cite{qiskit_contributors_qiskit_2023}, OpenFermion\cite{mcclean_openfermion_2020} and PauliArray on an Apple M2 MacBook were measured. Each mapping was repeated a hundred times and the average mapping times are reproduced on figure~\ref{fig:hamiltonian_mapping_plot}.

\begin{table*}
    \centering
    \caption{Molecular systems used to test the PauliArray functionalities}
    \begin{tabular}{c||c|c|c|c|c}
        \multirow{2}{*}{Molecule} & \multirow{2}{*}{Orbitals} & Pauli strings  & Single      & Double      & Total Pauli strings \\
                                  &                           & in Hamiltonian & excitations & excitations & in excitations      \\ \hline
        $\text{LiH}$              & 12                        & 631            & 16          & 76          & 640                 \\
        $\text{H}_2\text{O}$      & 14                        & 1086           & 20          & 120         & 1000                \\
        $\text{N}_2$              & 20                        & 2951           & 30          & 285         & 2340                \\
        $\text{NH}_3$             & 16                        & 3609           & 42          & 567         & 4620                \\
        $\text{C}_2\text{H}_2$    & 24                        & 6401           & 70          & 1645        & 13300               \\
        $\text{C}_2\text{H}_4$    & 28                        & 8919           & 96          & 3144        & 25344
    \end{tabular}
    \label{table:molecules}
\end{table*}

Since all three implementations use the same sparse representation of the fermionic Hamiltonian coefficient, the time complexities with the number of Pauli strings are similar. However, PauliArray being able to handle arrays of operators it allows to compute the operators in~\eqref{eq:update_flip_op_arrays_one} and~\eqref{eq:update_flip_op_arrays_two} to perform the Jordan-Wigner mapping faster than Qiskit ($\sim 6\times$) and OpenFermion ($\sim 3\times$).





\subsection{Application to the Adaptive Derivative Assembled Pseudo-Trotter Algorithm}

In the context of the variational quantum eigensolver algorithm, the ADAPT approach \cite{grimsley_adaptive_2019,tang_qubit-adapt-vqe_2021} is used to iteratively construct an ansatz circuit to prepare the ground state of a given Hamiltonian
\begin{align*}
    \hat{H} = \sum_i h_i \hat{P}_i
    .
\end{align*}

At step $k$ of this procedure a unitary transformation, the evolution of the Hermitian operator $\hat{A}_{R_k}$, is applied on the previous step circuit
\begin{align*}
    |\psi^{(k)}(\theta_1,\ldots,\theta_{k-1},\theta_{k})\rangle = e^{-i\theta_k\hat{A}_{R_k}}  |\psi^{(k-1)}(\theta_1,\ldots,\theta_{k-1})\rangle.
\end{align*}
The operator $\hat{A}_{R_k}$ is selected from a pool of Hermitian operators $\{\hat{A}_r\}$. The selected operator at step $k$, identified by its index $R_k$ is such that the variation in energy it produces from the last iteration state
\begin{align*}
    \frac{\partial}{\partial \theta_k}\langle \hat{H}\rangle_{\psi^{(k)}} = i \langle \psi^{(k-1)} | [\hat{H},\hat{A}_r] \psi^{(k-1)} \rangle
\end{align*}
is maximal in amplitude. This procedure requires to compute many commutators and to estimate their eigenvalues; two tasks where PauliArray can be used.

First, the pool of operators can be translated into an array of Pauli strings, when the pool contains single Pauli strings, or an operator array of (type 1 or type 2) for more general operators. In all cases, the commutator with the Hamiltonian involves the commutators between its Pauli strings and the Pauli strings of the operator pool. For example, with an operator pool of the form
\begin{align*}
    \hat{A}_r = \sum_j w_{rj} \hat{P}^\prime_{rj}
\end{align*}
the commutators to be estimated are
\begin{align}
    [\hat{H},\hat{A}_r] = \sum_{i,j} h_i w_{rj} [\hat{P}_i,\hat{P}^\prime_{rj}].
    \label{eq:adapt_commutators}
\end{align}

\begin{figure*}
    \centering
    \subfloat[Total time to compute the commutators between the molecular Hamiltonian and the excitation operators as a function of the total number of single Pauli commutators.\label{subfig:commutator_time_plot}]{\includegraphics{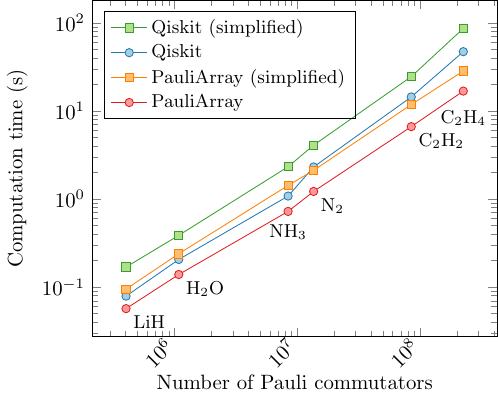}}
    \hfill
    \subfloat[The total number of Pauli strings in the commutators as computed by Qiskit and PauliArray as a function of the remaining number of Pauli strings after simplification. The distance to the dashed line (slope 1) represents the Pauli strings removed by the simplification. \label{subfig:commutator_tot_num_terms_plot}]{\includegraphics{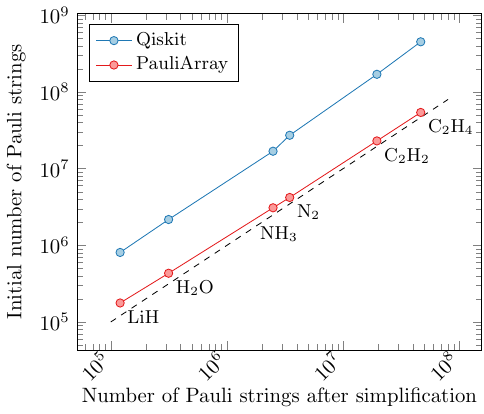}}
    \caption{Computation of the commutators between the Hamiltonian and the single and double excitations for various small molecules.}
    \label{fig:key_operations}
\end{figure*}

The performance of PauliArray to compute these commutators was assessed considering the same previously used set of small molecules (see table~\ref{table:molecules}). For each molecule, a pool containing the single and double fermionic excitations is built and the Jordan-Wigner mapping is used to convert them into Operators of respectively 2 and 8 Pauli strings. Then the time it takes for the typical implementations from Qiskit and PauliArray to compute the commutators between the Hamiltonian and all the excitation operators is measured. These tests were again conducted on an Apple M2 MacBook and repeated a hundred times to ensure consistency.

The average computation time is shown on figure~\ref{subfig:commutator_time_plot} as a function of the total number of single commutators $[\hat{P}_i,\hat{P}^\prime_{rj}]$ (given by the product between the number of Pauli strings in the Hamiltonian with the total number of Pauli strings in all the excitation operators). In addition, the figure also shows the times when a simplification step (section~\ref{ssec:simplification}) is also performed.

While the time complexities are all similar, PauliArray offer a sizeable advantage. Overall, PauliArray is roughly twice as fast as Qiskit to accomplish similar tasks. The main difference between the two implementations is that Qiskit computes commutators as the difference between two compositions ($\hat{P}_1 \hat{P}_2 - \hat{P}_2 \hat{P}_1$) while PauliArray uses a single composition as discussed in section~\ref{ssec:commutation}.

One other consequence of this approach is that the number of Pauli strings involved in the commutators before simplification is significantly smaller as shown on figure~\ref{subfig:commutator_tot_num_terms_plot}. By identifying the non-vanishing commutators beforehand, PauliArray avoids to compute most of the Pauli strings which would be removed by the simplification. Consequently, the simplification process should be easier and faster for PauliArray. However, Qiskit offers a better optimized method to that end, which limits the advantage of PauliArray.

\subsection{Expectation Values Estimation}
\label{ssec:estimation}

Another important application of decomposing operators and observables on the basis of Pauli strings is to provide a way to estimate their expectation values and their variances. One of the central ideas in PauliArray in this context is that the expectation value of an array of operators or Pauli strings for given a state $|\psi\rangle$ is an array of expectation values with the same dimensions
\begin{align*}
    \mathbb{E}_\psi(O_{i})  = \langle\psi|\hat{O}_{i}|\psi\rangle.
\end{align*}
Similarly, the covariance between two arrays of operators is an array combining dimensions of the two arrays
\begin{align*}
    \text{Covar}_\psi(O^{(1)}_i,O^{(2)}_j)
     & = \langle\psi|\hat{O}^{(1)}_{i}\hat{O}^{(2)}_{j}|\psi\rangle                               \\
     & \quad- \langle\psi|\hat{O}^{(1)}_{i}|\psi\rangle\langle\psi|\hat{O}^{(2)}_{j}|\psi\rangle.
\end{align*}

For all data structures in PauliArray these quantities can be directly computed by providing the expectation values and the covariances of the underlying array of Pauli strings. For example, the expectation values for a type 1 operator array
\begin{align}
    \hat{O}_{i} = \sum_s w_{is} \hat{P}_{is}
    \label{eq:exemple_operator_array}
\end{align}
are given by
\begin{align*}
    \mathbb{E}_\psi(O_{i}) = \sum_s w_{is}\, \mathbb{E}_\psi(P_{is})
\end{align*}
while its covariance array is
\begin{align*}
    \text{Covar}_\psi(O_i,O_j) = \sum_{st} w_{is}w_{jt}\, \text{Covar}_\psi(P_{is},P_{jt}).
\end{align*}

This interface offers the flexibility to use other libraries to carry out the estimation of the expectation values and covariance matrix of the underlying array of Pauli strings. However, PauliArray also provide functionalities to that end, using established techniques to minimize resources needed to do so. The first step of this process is to partition a set of Pauli strings into subsets of commuting Pauli strings so that they share the same eigenbasis and therefore can be estimated using the same measurement circuit. The next step is to find the transformation and construct the corresponding quantum circuit which enables this measurement to be carried in the computational basis. The two following sections (\ref{ssec:partition} and \ref{ssec:diagonalization}) illustrate how PauliArray can be used to these purposes.

\subsection{Commuting Pauli Partitioning}
\label{ssec:partition}

Partitioning a set of Pauli strings into subsets of commuting Pauli strings is an application related to the measurement of observables~\cite{gokhale_minimizing_2019, yen_measuring_2020} but also useful to Hamiltonian evolution~\cite{martinez-martinez_assessment_2022}. This problem is usually approached by constructing an undirected graph where each node represents one of the Pauli strings involved in the Hamiltonian or in the observable. A pair of nodes associated to a pair of commuting Pauli strings are then connected by an edge. The commutation graph for a set of Pauli strings $\{\hat{P}_i\}$ can be constructed via its adjacency matrix whose elements are given by
\begin{align*}
    A_{ij} = \mathbf{z}_i \cdot \mathbf{x}_j + \mathbf{x}_i \cdot \mathbf{z}_j + 1 \pmod{2}
\end{align*}
such that $A_{ij} = 1$ iff $[\hat{P}_i,\hat{P}_j] = 0$. Using broadcasting within PauliArray, the adjacency matrix can be constructed with a single line of code.

Finding the minimal number of subsets of commuting Pauli strings that cover the whole graph is then equivalent to the Minimum Clique Cover problem. Many heuristic solutions to this problem exists and their performances for this particular instance have been extensively studied \cite{verteletskyi_measurement_2020} and won't be discussed here. The result of these solutions are sets of indices identifying the Pauli strings in the commuting subsets. These subsets can then be extracted using the indexing functionalities of PauliArray.

\subsection{Diagonalization}
\label{ssec:diagonalization}

By definition, commuting Pauli strings share a common eigenbasis and therefore can be simultaneously diagonalized using an operator $\hat{C}$ made only of Clifford operations\cite{yen_measuring_2020}
\begin{align*}
    \hat{C} \hat{P}_i \hat{C}^\dagger = \hat{D}_i \quad \text{for}\quad [\hat{P}_i,\hat{P}_j] = 0
\end{align*}
where a diagonal Pauli string is a tensor product of only diagonal operators
\begin{align*}
    \hat{D}        & = \bigotimes_{q=0}^{n-1} \hat{\sigma}_q
                   &                                         & \text{with} &
    \hat{\sigma}_q & \in \left\{\hat{Z}, \hat{I}\right\}.
\end{align*}

The diagonalization submodule of PauliArray implements the simultaneous diagonalization algorithm described in \cite{berg_circuit_2020}. Once a set of commuting Pauli strings has been identified and encoded as a one-dimensional array of Pauli strings, this submodule can be used to construct the diagonalization operator $\hat{C}$ as a quantum circuit. The array of diagonalized Pauli strings can easily be obtained to be used in the target application.

Since the conjugation by Clifford operations is available in PauliArray, it can also be used to experiment with new diagonalization algorithms.

\section{Conclusion}

The PauliArray library defines a few data structures to represent numerous Pauli strings and operators decomposed on the basis of Pauli strings. It provides a wide range of functionalities to transform and combine these data structures. In particular, the use of broadcasting simplifies some otherwise cumbersome operations.

This paper explores various use cases where the functionalities of PauliArray can be applied to simplify some procedures or to improve their efficiency. In particular, it has been shown that PauliArray compares favourably to other libraries for the fermion-to-qubit mapping and the computation of commutators. This tool will likely find uses elsewhere since the manipulation of Pauli strings is ubiquitous to quantum computing and to many fields of theoretical and computational physics. While the multidimensionality aspect of PauliArray will remain, its implementation might evolve in the future to improve performance. In particular, the development of a dedicated backend to handle arrays of bits could prove beneficial to its overall performance.

\section*{Acknowledgment}

The authors would like to thank Alexandre Foley, Camille Le Calonnec, Olivier Nahman-L\'evesque and Jean-Fr\'ed\'eric Laprade for fruitful discussions and relevant feedback.

\appendix

\label{app:one_two_op_exp}

Combining~\eqref{eq:creation_annihilation_operators} with itself for two different indices yields
\begin{align*}
    \hat{a}_i^\dagger \hat{a}_j
     & =
    \frac{1}{4}
    \hat{X}_q^{M_{qi}}
    \hat{Z}_q^{\theta_{ip}[\mathsf{M}^{-1}]_{pq}}
    \left(
    1 + \hat{Z}_q^{ [\mathsf{M}^{-1}]_{iq} }
    \right)
    \\
     & \quad
    \hat{X}_q^{M_{qj}}
    \hat{Z}_q^{\theta_{jp}[\mathsf{M}^{-1}]_{pq}}
    \left(
    1 - \hat{Z}_q^{ [\mathsf{M}^{-1}]_{jq} }
    \right)   .
\end{align*}
All the $\hat{X}$ operators can be moved to the front of the expression using the following commutation relations
\begin{align*}
    \hat{Z}_q^{\theta_{ip}[\mathsf{M}^{-1}]_{pq}}
    \hat{X}_q^{M_{qj}}
     & =
    (-1)^{\theta_{ij}}
    \hat{X}_q^{M_{qj}}
    \hat{Z}_q^{\theta_{ip}[\mathsf{M}^{-1}]_{pq}} \\
    \hat{Z}_q^{ [\mathsf{M}^{-1}]_{iq} } \hat{X}_q^{M_{qj}}
     & =
    (-1)^{\delta_{ij}}\hat{X}_q^{M_{qj}}\hat{Z}_q^{ [\mathsf{M}^{-1}]_{iq} }.
\end{align*}
Regrouping the associated $\hat{Z}$ operators leads to the expression
\begin{align*}
    \hat{a}_i^\dagger \hat{a}_j
     & =
    \frac{1}{4}
    \hat{X}_q^{M_{qi} + M_{qj}}
    \hat{Z}_q^{(\theta_{ip} + \theta_{jp})[\mathsf{M}^{-1}]_{pq}}
    \\
     & \quad
    \left(
    1 + (-1)^{\delta_{ij}}\hat{Z}_q^{ [\mathsf{M}^{-1}]_{iq} }
    \right)
    \left(
    1 - \hat{Z}_q^{ [\mathsf{M}^{-1}]_{jq} }
    \right)
\end{align*}
which can be rewritten as the composition of the three operators (see~\eqref{eq:one_body_operator_paulis})
\begin{align}
    \begin{split}
        \hat{U}^{(2)}_{ij}  & = (-1)^{\theta_{ij}} \hat{X}_q^{M_{qi} + M_{qj}}
        \hat{Z}_q^{(\theta_{ip} + \theta_{jp})[\mathsf{M}^{-1}]_{pq}}
        \\
        \hat{F}^{(2+)}_{ij} & = 1 + (-1)^{\delta_{ij}} \hat{Z}_q^{ [\mathsf{M}^{-1}]_{iq} }
        \\
        \hat{F}^{(1-)}_{i}  & = 1 - \hat{Z}_q^{ [\mathsf{M}^{-1}]_{iq} }.
    \end{split}
    \label{eq:update_flip_op_arrays_one}
\end{align}
Equation~\eqref{eq:two_body_operator_paulis} is obtained in a similar way with the combination of~\eqref{eq:creation_annihilation_operators} with four different indices. The definition of the operator arrays are the following
\begin{align}
    \begin{split}
        \hat{U}^{(4)}_{ijkl}  & = (-1)^{\theta_{ij} + \theta_{ik} + \theta_{il} + \theta_{jk}  + \theta_{jl} + \theta_{kl}} \\&\quad\hat{X}_q^{M_{qi} + M_{qj} + M_{qk} + M_{ql}}
        \hat{Z}_q^{(\theta_{ip} + \theta_{jp} + \theta_{kp} + \theta_{lp})[\mathsf{M}^{-1}]_{pq}}
        \\
        \hat{F}^{(4+)}_{ijkl} & = 1 + (-1)^{\delta_{ij} + \delta_{ik} + \delta_{il}} \hat{Z}_q^{ [\mathsf{M}^{-1}]_{iq} }
        \\
        \hat{F}^{(3+)}_{ijk}  & = 1 + (-1)^{\delta_{ij} + \delta_{ik} } \hat{Z}_q^{ [\mathsf{M}^{-1}]_{iq} }
        \\
        \hat{F}^{(2-)}_{ij}   & = 1 - (-1)^{\delta_{ij}} \hat{Z}_q^{ [\mathsf{M}^{-1}]_{iq} }
        \\
        \hat{F}^{(1-)}_{i}    & = 1 - \hat{Z}_q^{ [\mathsf{M}^{-1}]_{iq} }.
    \end{split}
    \label{eq:update_flip_op_arrays_two}
\end{align}

\label{app:other}

\bibliographystyle{IEEEtran}
\bibliography{biblio.bib}

\end{document}